\documentclass[]{article}
\def\beq{\begin{eqnarray}}
\def\eeq{\end{eqnarray}}
\begin{document}
\title{\bf Black hole state counting in loop quantum gravity}
\author{P. Mitra\thanks{parthasarathi.mitra@saha.ac.in}\\
Saha Institute of Nuclear Physics,\\
1/AF Bidhannagar, Calcutta 700064}
\date{}
\maketitle
\begin{abstract}
Counting of microscopic states of black holes is discussed within the framework of
loop quantum gravity. There are two different ways, one 
allowing for all spin states and the other involving only pure horizon states.
The number of states with a definite value of the total spin is also found.  
\end{abstract}
\bigskip 
\bigskip

\section{Introduction}
The framework of quantum gravity 
known as loop quantum gravity has been able to yield a detailed counting of
microscopic quantum states corresponding to a black hole.
A start was made in
\cite{ash} in the direction of quantizing a black hole 
characterized by an
isolated horizon. The quantum states arise
when the cross sections of the horizon
are punctured by spin networks. The spin quantum numbers
$j,m$, which characterize the punctures, then label the quantum states. The entropy
is obtained by counting the possibilities of such 
labels that are consistent with a fixed area of the cross section
\cite{ash}.

A calculation of the entropy was carried out in \cite{meissner} using
a recursion relation technique. In \cite{gm}, 
(see also \cite{cor} on this issue) a combinatorial
method was introduced, which in addition to counting states also gives the dominant
configuration of spins, namely the configuration yielding the maximum number of
states. However, the two calculations give different results. 
The difference is due to the fact that
while \cite{meissner} takes into account only the spin projection ($m$) labels of
the microstates, thus counting what may be called the pure {\em horizon} states, 
\cite{gm} and \cite{cor} take into
account the spin $j$, which is relevant for the eigenvalues of the area operator, as well as
the $m$-labels. There are two constraints to be satisfied. While one of them,
the spin projection constraint, can be expressed solely in
terms of the $m$-labels, the other constraint involving the area of the horizon,
explicitly uses the $j$-labels.

\section{Counting of states}
We temporarily use units such that $4\pi\gamma\ell_P^2=1$, where $\gamma$ is the 
so-called Barbero-Immirzi parameter involved in the quantization
and $\ell_P$ the Planck length. Setting the classical area $A$ of the
horizon equal to the eigenvalue (for a specific spin configuration of punctures on the
horizon) of the area operator we find
\beq A=2\sum_{j,m}s_{j,m}\sqrt{j(j+1)},\label{areacon2}\eeq
where $s_{j,m}$ is the number of punctures carrying spin quantum numbers $j,m$. 
Such a spin configuration will be admissible if it obeys
(\ref{areacon2}) together with the {\em spin projection constraint}
\beq 0=\sum_{j,m}ms_{j,m}\;.\label{newc}\eeq

The total number of quantum states for these configurations
is
\beq d_{s_{j,m}}={(\sum_{j,m} s_{j,m})!\over\prod_{j,m}s_{j,m}!}\;.\label{newco}\eeq
To obtain the dominant permissible configuration that contributes the largest number
of quantum states, we maximize $\ln d_{s_{j,m}}$ by varying $s_{j,m}$ subject to
the constraints using Stirling's approximation: 
\beq\ln d_{s_{j,m}}&=&(\sum_{j,m} s_{j,m})\ln \sum_{j,m}s_{j,m}-
\sum_{j,m} (s_{j,m}\ln s_{j,m}),\nonumber\\
\delta\ln d_{s_{j,m}}&=&(\sum_{j,m}\delta s_{j,m})\ln \sum_{j,m}s_{j,m}
-\sum_{j,m}(\delta s_{j,m}\ln s_{j,m}).\eeq
The condition for the maximum can be expressed in terms of two Lagrange multipliers
$\lambda,\alpha$:
\beq \ln s_{j,m}-\ln \sum_{j,m}s_{j,m}=-2\lambda\sqrt{j(j+1)}-\alpha m,\eeq
whence
\beq {s_{j,m}\over\sum s_{j,m}}=e^{-2\lambda\sqrt{j(j+1)}-\alpha
m}\;.\label{news}\eeq
Consistency requires that $\lambda$ and $\alpha$ be related to each other by 
\beq\sum_j
e^{-2\lambda\sqrt{j(j+1)}}\sum_me^{-\alpha m}=1.\eeq 
In order that (\ref{news}) satisfies
the spin projection constraint, we need 
$\sum_mme^{-\alpha m}=0$ for each $j$, which essentially implies $\alpha=0$. 
Therefore, the consistency condition becomes
\beq\sum_{j,m}e^{-2\lambda\sqrt{j(j+1)}}=1.\eeq
Numerical solution of this equation yields $\lambda=0.861.$ 
Note that each 
$s_{j,m}$ is proportional to the area $A$ because of the area constraint.
Further, in general,
\beq \ln d_{s_{j,m}}=\lambda A+\alpha \sum_{j,m}s_{j,m}m,\eeq
in which the last term vanishes in the present situation 
because of the spin projection condition, but will appear later.

The total number of quantum states for all permissible configurations is clearly
$d=\sum_{ s_{j,m}}d_{s_{j,m}}$. To estimate $d$ we expand $\ln d$ around the
dominant configuration (\ref{news}), which we shall denote by $\bar s_{j,m}$. Thus
\beq\ln d=\ln d_{\bar s_{j,m}}-{1\over 2}\sum\delta s_{j,m}K_{j,m;j'm'}\delta
s_{j'm'}+o(\delta s_{j,m}^2)\eeq 
where $\delta s_{j,m}= s_{j,m}-\bar s_{j,m}$ and $K$ is the symmetric matrix
\beq K_{j,m;j'm'}=\delta_{jj'}\delta_{mm'}/\bar s_{j,m}-1/\sum_{k,l}\bar s_{k,l}.\eeq 
The sum over each $\delta s_{j,m}$ can be approximated by a Gaussian integral. The 
eigenvalues of $K$ are proportional to $1/A$, so each integration produces a
factor $\sqrt A$. The number of these factors is two less than the number of
$s_{j,m}$ because of the two constraints on the $\delta s_{j,m}$. On the other hand, we see from
(\ref{newco}) that the combinatorial number contains one  $\sqrt A$ for
each $s_{j,m}$ in the denominator and one more in the numerator because 
\beq n!\approx\sqrt{2\pi n}(\frac{n}{e})^n.\eeq
In all, one factor of  $\sqrt A$ survives in the denominator, so that
\beq d={{\rm constant}\over\sqrt A}\;e^{\lambda A}\;,\eeq
leading to the formula \cite{gm}
\beq S=\lambda {A\over 4\pi\gamma\ell_P^2}-\frac12 \ln {A\over 4\pi\gamma\ell_P^2}
\label{s}\eeq
for entropy. The origin of the $\sqrt A$ in $d$ or $\frac12\ln A$ in $\ln d$ can be
easily traced in this approach: it is the condition $\sum ms_{j,m}=0$.

\section{Counting of horizon states}
The above calculation assumed that $j$ is a relevant quantum
number. An alternative procedure \cite{ash, meissner}, is to count
the states of the horizon Hilbert space alone. Here, following \cite{gm2}, we consider
the number $s_m$ of punctures carrying spin projection $m$, ignoring what spins $j$
they are associated with. Clearly,
\beq s_m=\sum_js_{j,m},\quad j=|m|,|m|+1,|m|+2,... .\eeq
For the $s_m$ configuration the number of states is
$d_{s_m}=(\sum_{m}s_{m})!/\prod_{m}s_{m}!$ and the total number of states is
obtained by summing over all configurations. As in the earlier case, the sum can be
approximated by maximizing $\ln d_{s_m}$ subject to the two conditions.
The constrained extremization conditions for
variation of $s_{j,m}$ are
\beq -\big[\ln{s_{m}\over\sum_ms_{m}}+2\lambda\sqrt{j(j+1)}+\alpha m\big]
=0.\label{consm}\eeq
All these equations cannot hold for arbitrary $j$ even for a fixed $m$,
because inconsistencies will arise for nonzero $\lambda$. In fact, for any fixed $m$
the above equality can be valid for at most one $j$ -- say $j(m)$.
For $j\neq j(m)$, the first derivative becomes nonzero. Such a situation can arise
if and only if $\ln d_{s_m}$ is maximized at the boundary (in the space of all
permissible configurations) for all $j\neq j(m)$ and at an interior point for
$j=j(m)$. This means that for the dominant configuration, $s_{j,m}=0$ for all $j\neq
j(m)$: the corresponding first derivative is then only required to be zero or
negative because in any variation $s_{j,m}$ can only increase from its zero value.
Thus, $s_m=s_{j(m),m}$ for the dominant configuration
and further, for $\lambda>0$, $j(m)=j_{\rm min}(m)$, the minimum value for the $m$.
For $m\neq 0$, we have $j_{\rm min}(m)=|m|$. 

Then (\ref{consm}) gives
\beq {s_{m}\over\sum_{m}s_{m}}=e^{-2\lambda\sqrt{j_{\rm min}(m)(j_{\rm min}(m)+1)}-\alpha
m}\label{consf}\;.\eeq
As before, $\alpha=0$ because of the spin
projection constraint. 

The configuration (\ref{consf}) implies that
the entropy is given by (\ref{s}) in terms of $\lambda$, which is now determined
by the altered consistency relation
\beq 1=\sum_{j}2e^{-2\lambda\sqrt{j(j+1)}}.\eeq
%
Note that for $\lambda$ zero or negative, such relations would be impossible to
satisfy, hence no such solutions exist.

This equation for $\lambda$ aagrees with that of \cite{meissner}.
%
%

\section{Vanishing total spin projection?}
We have imposed the condition of vanishing spin projection in the above calculation.
It is interesting to fix the total spin projection to some value and
see how the number of states changes with this quantity. Thus we set
\beq \sum_{j,m}ms_{j,m}=p.\eeq
The main difference with earlier equations will arise from the
fact that $\alpha$ will no longer vanish. Let us introduce
\beq F(\lambda,\alpha)\equiv\sum_{j,m}e^{-2\lambda\sqrt{j(j+1)}-\alpha m}.\eeq
Then we have the conditions
\beq F(\lambda,\alpha)&=&1,\nonumber\\
\frac{p}{A}&=&\frac{{\partial F\over\partial\alpha}}{{\partial F\over\partial\lambda}}.
\eeq
These two equations determine $\lambda,\alpha$ in terms of $\frac{p}{A}$.
On the basis of what we already know, we can write the general equation
\beq \ln d=\lambda A+\alpha p-\frac12\ln A.\eeq
Now 
\beq \lambda A+\alpha p=A( \lambda+\alpha\frac{{\partial F\over\partial\alpha}}
{{\partial F\over\partial\lambda}})=A(\lambda(\alpha)-\alpha{d\lambda\over d\alpha}),\eeq
where $\lambda(\alpha)$ is understood to be the solution of $F=1$.
If $\frac{p}{A}$ is small, $\alpha$ can be taken to be small, and by
Taylor expansion of $\lambda(\alpha)$ about $\alpha=0$, we find
\beq \lambda A+\alpha p\approx A(\lambda(0)-\frac{\alpha^2}{2}{d^2\lambda\over d\alpha^2}|_{\alpha=0}).\eeq
Note that
\beq {d^2\lambda\over d\alpha^2}|_{\alpha=0}=
-{{\partial^2 F\over\partial\alpha^2}
\over{\partial F\over\partial\lambda}}|_{\alpha=0} =
\frac{\sum_{j,m}m^2e^{-2\lambda\sqrt{j(j+1)}}}
{2\sum_{j,m}\sqrt{j(j+1)}e^{-2\lambda\sqrt{j(j+1)}}}=k,\eeq
say, which is positive.
Again, by expanding in $\alpha$ for small $\alpha$, we find
\beq\frac{p}{A}=\alpha
{{\partial^2 F\over\partial\alpha^2}
\over{\partial F\over\partial\lambda}}|_{\alpha=0} =-\alpha k.\eeq
Hence, 
\beq\lambda A+\alpha p=A\lambda(0)-\frac{p^2}{2kA},\eeq
and
\beq \ln d=\lambda(0) A-\frac{p^2}{2kA}-\frac12\ln A.\eeq
Note that $\lambda(0)$ here is the same as the $\lambda$
of the earlier situation where $\alpha=0$.

The number of states for a definite value of $p$ is thus (cf. \cite{meissner})
\beq d(p)\sim{\exp(\lambda(0) A-\frac{p^2}{2kA})\over\sqrt{A}}
\approx{\exp(\lambda(0) A)(1-\frac{p^2}{2kA})\over\sqrt{A}}.\eeq
The number of states for total spin $J$ and the same spin projection
is then
\beq d(J)-d(J+1)\sim(2J+1)
{\exp(\lambda(0) A)\over2kA\sqrt{A}}.\eeq
The number of states for (small) total spin $J$ is 
\beq N(J)\sim(2J+1)^2
{\exp(\lambda(0) A)\over2kA\sqrt{A}}.\eeq
In particular, for $J=0$, this becomes
\beq N(0)\sim{\exp(\lambda(0) A)\over2kA^{3/2}}.\eeq
These may be compared with the results found in \cite{kaul}.
The reason for the disagreement between the
coefficient of the log correction in the entropy
found there and that in \cite{gm0,meissner,gm,cor}
is seen to be that the {\it total spin} is involved in the former,
while in the latter, following the loop quantum gravity
literature, only the spin projection constraint is imposed.

\section*{Post script}
However, it has now been suggested \cite{perez} that in an alternative
quantization, at least for large area, the total
spin $J$ has to vanish.
\section*{Acknowledgments}
This talk was based on work done in collaboration with
Amit Ghosh. The calculation involving non-zero spin projection 
was done following discussions with Romesh Kaul.


\end{document}